\documentclass[10pt,twoside]{hsqcd}
\usepackage{epsf,amsmath}
\usepackage{graphicx}

\newcommand{\bo}[1]{\mbox{\boldmath $ #1 $}}
\begin{document}

\title{
EXPLICIT EQUATIONS FOR RENORMALIZATION PRESCRIPTIONS
IN THE CASE OF PION-NUCLEON SCATTERING
\thanks{
This work is supported by INTAS (project 587, 2000), Ministry of
Education of Russia (Programme ``Universities of Russia'') and L.
Meltzers H\o yskolefond (Studentprosjektstipend, 2004).
}}

\author{
A.VERESHAGIN\\
University of Bergen and St.~Petersburg State University\\
E-mail: Alexander.Vereshagin@ift.uib.no
}

\maketitle

\begin{abstract}
\noindent
This talk is a natural continuation of that given by V.~Vereshagin
\cite{vvv-hsqcd}.
We discuss some details not covered in that talk and review the
calculational technique using the
$\pi N$
elastic scattering as an example. Finally, we briefly mention some
results of numerical comparison with experimental data. More technical
details can be found in the talk by K.~Semenov-Tian-Shansky
\cite{kstsh-hsqcd}
devoted to the analysis of elastic
$KN$
scattering.
\end{abstract}

%%%%%%%%%%%%%%%%%%%%%%%%%%%%%%%%%%%%%%%%%%%%%%%%%%%%%%%%%%%%%%%%%%%%%%
\markboth{\large \sl
A.~VERESHAGIN\\
\hspace*{2cm} HSQCD 2004}
{\large \sl \hspace*{1cm}
EXPLICIT EQUATIONS FOR RENORMALIZATION ...}

%%%%%%%%%%%%%%%%%%%%%%%%%%%%%%%%%%%%%%%%%%%%%%%%%%%%%%%%%%%%%%%%%%%%%%
\section{General notes}

As explained in the talk
\cite{vvv-hsqcd}, our approach does not assume any
``nuclear democracy''.
In contrast, it discriminates between stable particles and resonances.
Only stable particles survive as asymptotic states, and it is the
stable sector where the
$S$-matrix
is unitary (see, e.g.
\cite{Veltman}).
If we restrict ourselves by a consideration of the strong non-strange
sector, then the only stable particles are pions and nucleons. Hence,
to illustrate the application of our technique by the relatively
simple process, we can choose among
$\pi\pi$, $NN$,
and
$\pi N$-elastic
scattering (along with the cross-symmetric processes). Our choice of
($\pi N$)
is mainly dictated by the absence of extra phenomenological symmetries
appearing in the former two reactions and, at the same
time, by the relatively rich set of experimental data.

When working in the framework of effective theory one has to take
account of all possible vertices and resonances which can contribute
to the amplitude of the reaction under consideration. Since the
perturbation theory which we rely upon is of Dyson's type, we need to
construct the perturbation series order by order, starting from the
tree level. However, at this very first step we immediately meet the
difficulty because to obtain the tree level amplitude we need to sum
an infinite number of contact vertices and exchange graphs
(Fig.~\ref{fig:1}).

%%%%%%%%%%%%%%%%%%%%%%%%%%%%%%%%%%%%%%%%%%%%%%%%%%%%%%%%%%%%%%%%%%%%%%
\begin{figure}[ht]
\begin{center}
%%%%%%%%%%%%%%%%%%%%%%%%% tree diagrams
\begin{picture}(350,20)(25,-10)
\put(0,0){
\begin{picture}(100,20)(0,0) % contact
\put(0,-5){\shortstack{$\displaystyle\sum_{ \rm vertices \atop
\mbox{} }^{\infty}$}}
\put(40,0){\circle*{3}}
\put(40,0){\line(-1,1){10}}
\put(40,0){\line(-1,-1){10}}
\put(30,-10){\vector(1,1){7}} % spinor arrow
\put(40,0){\line(1,-1){10}}
\put(40,0){\vector(1,-1){7}} % spinor arrow
\put(40,0){\line(1,1){10}}
\put(60,-10){\shortstack{\boldmath$,$}}
\end{picture}
}

\put(100,0){
\begin{picture}(100,20)(0,0) % "s"
\put(0,-5){\shortstack{$\displaystyle\sum_{ \rm vertices, \atop
\rm resonances }^{\infty}$}}
\put(40,0){\circle*{3}}
\put(40,0){\line(-1,1){10}}
\put(40,0){\line(-1,-1){10}}
\put(30,-10){\vector(1,1){7}} % spinor arrow
\multiput(40,0)(1,0){20}{\circle*{2}}  % propagator
\put(40,0.5){\vector(1,0){13}} % spinor arrow
\put(40,-0.5){\vector(1,0){13}} % spinor arrow
\put(45,-13){\shortstack{$R_s$}}
\put(60,0){\circle*{3}}
\put(60,0){\line(1,1){10}}
\put(60,0){\line(1,-1){10}}
\put(60,0){\vector(1,-1){7}} % spinor arrow
\put(80,-10){\shortstack{\boldmath$,$}}
\end{picture}
}

\put(200,0){
\begin{picture}(100,20)(0,0) % "t"
\put(0,-5){\shortstack{$\displaystyle\sum_{ \rm vertices, \atop
\rm resonances}^{\infty}$}}
\put(50,-10){\circle*{3}}
\put(50,-10){\line(-1,-1){10}}
\put(40,-20){\vector(1,1){7}} % spinor arrow
\put(50,-10){\line(1,-1){10}}
\put(50,-10){\vector(1,-1){7}} % spinor arrow
\multiput(50,-10)(0,1){20}{\circle*{2}}  % propagator
\put(53,-5){\shortstack{$R_t$}}
\put(50,10){\circle*{3}}
\put(50,10){\line(-1,1){10}}
\put(50,10){\line(1,1){10}}
\put(75,-10){\shortstack{\boldmath$,$}}
\end{picture}
}

\put(300,0){
\begin{picture}(100,20)(0,0)  % "u"
\put(0,-5){\shortstack{$\displaystyle\sum_{ \rm vertices, \atop
\rm resonances}^{\infty}$}}
\put(40,0){\circle*{3}}
\multiput(40,0)(1,0){20}{\circle*{2}}  % propagator
\put(40,0.5){\vector(1,0){13}} % spinor arrow
\put(40,-0.5){\vector(1,0){13}} % spinor arrow
\put(45,-13){\shortstack{$R_u$}}
\put(60,0){\line(1,-1){10}}
\put(60,0){\vector(1,-1){7}} % spinor arrow
\put(60,0){\circle*{3}}
\put(30,-10){\line(1,1){15}}
\put(30,-10){\vector(1,1){7}} % spinor arrow
\put(55,15){\line(1,1){10}}
\put(60,0){\line(-1,1){25}}
\put(55,5){\oval(20,20)[tl]}
\put(80,-10){\shortstack{\boldmath$.$}}
\end{picture} }

\end{picture}
\end{center}
\caption{Tree graphs:
$R_s$, $R_t$ and $R_u$ stand for all admissible resonances in the
$s$-, $t$-, and $u$-channels, respectively; summation over all
possible kinds of vertices is implied, though the summation order is
still unspecified.
\label{fig:1}}
\end{figure}
%%%%%%%%%%%%%%%%%%%%%%%%%%%%%%%%%%%%%%%%%%%%%%%%%%%%%%%%%%%%%%%%%%%%%%

\noindent
The resulting sum is nothing but functional series, thus the problem
of summation order is essential one. As it is demonstrated in
\cite{vvv-hsqcd,AVVV1,POMI,AVVV2},
our approach gives a way to overcome the obstacle. Simply speaking,
the recipe we suggest reads:
\begin{enumerate}
\item
Classifying all possible graphs and switching to the
{\em minimal parametrization}
\cite{AVVV2}
single out the set of
{\em resultant}
parameters of the given level (here --- tree level). The latter are
assigned the
{\em physical}
values with the help of relevant renormalization prescriptions (RP's).
\item
Being guided by the
{\em uniformity}
and
{\em summability}
\cite{vvv-hsqcd}
principles use the Cauchy formula for given order (tree level)
amplitude in certain domain of the space of kinematical variables.
\item
Equating different expressions for the amplitude
(the latter results from the Cauchy formula application) in the
domains of their mutual validity, obtain the system of
{\em bootstrap}
equations. The latter allow one to specify the exact expressions of
the amplitude under consideration and give restrictions for the values
of
{\em physical}
parameters of the theory.
\end{enumerate}

In this talk we shall take a closer look at the first and the last
steps.

%%%%%%%%%%%%%%%%%%%%%%%%%%%%%%%%%%%%%%%%%%%%%%%%%%%%%%%%%%%%%%%%%%%%%%
\section{Minimal (resultant) vertices and renormalization conditions}

As it is seen from
Fig~\ref{fig:1},
there are Hamiltonian%
\footnote{In
\cite{AVVV2}
it is explained why it is preferably to use the effective Hamiltonian,
rather than Lagrangian when constructing a theory with unlimited
number of field derivatives.}
three- and four-leg couplings and masses which parametrize the tree
level amplitude in our case.

Minimal parametrization is a first step toward the constructing of
so-called
{\em essential}
parameters
\cite{WeinMONO,AVVV2} ---
the
{\em independent}
parameters needed to describe the (on-shell) S-matrix. In case of
general process amplitude of arbitrary loop order the minimal
couplings are the natural building blocks for the resultant parameters
of which, in turn, the essential parameters can be constructed.
However, in case of triple vertices at tree level, this structure gets
simplified, and all the contributing three-leg minimal couplings
appear also to be ``resultant''.

The minimal vertices are, roughly speaking, the on-shell vertices. One
just needs to take the
{\em effective vertex}
of a given order (at tree level this is a matrix element of the sum of
all Hamiltonian vertices constructed of a given set of fields with all
possible derivatives and matrix structures), put it on the mass shell,
present the result in a Lorentz-covariant form and cross the wave
functions out. The structure surviving after this is done, being
considered as a function of independent components of
{\em off-shell} momenta%
\footnote{
Energy-momentum conservation is, of course, implied. For the precise
definition of minimal vertex and the related classification see
\cite{AVVV2}
}
is called the minimal vertex. The coefficients in the formal series
for the corresponding formfactors are called the minimal couplings%
\footnote{
They are, of course, functions of initial Hamiltonian couplings.
However the latter functional dependence is not of interest anymore:
we are not going to fix any of couplings in the initial Hamiltonian,
rather, we will prefer to operate with minimal (resultant) parameters
directly.
}.
One easily observes that the tree-level triple minimal couplings are
constants, because on the mass shell any triple vertex does not depend
upon external momenta. For example, all the minimal vertices with
resonances of isospin
$\frac{1}{2}$
and half-integer spin
$l+\frac{1}{2}$
contributing to our process at tree level can be listed as the
following ``Hamiltonian monomials''%
\footnote{
Lacking space here, we do not list the remaining vertices with
half-integer spin resonances and those with integer spin contributing
in
$t$-channel.
}:
\begin{description}
\item{}
$
g_{\widehat{R}}
\overline{N}\bo{\sigma}
\widehat{R}_{\mu_1\ldots\mu_l}
\partial^{\mu_1}\!\!\!\!\ldots\partial^{\mu_l}\bo{\pi}
+ H.c.\;
$
for the resonance parity
$P = (-1)^{l+1}$, and
\item{}
$
ig_{R}
\overline{N} \bo{\sigma} \gamma_5
R_{\mu_1\ldots\mu_l}
\partial^{\mu_1}\!\!\!\!\ldots\partial^{\mu_l}\bo{\pi}
+ H.c.\;
$
for the resonance parity
$P = (-1)^l$,
\end{description}
where
$\sigma_a$
stands for Pauli matrix,
$\pi$, $N$, and $R$
denote pion, nucleon and resonance fields, respectively, while
$g$'s
are the minimal coupling constants which, of course, depend on the
resonance spin and mass. The essence of the reduction theorem proved
in
\cite{AVVV2}
is that any vertex that differs from the listed above by the
number (or/and position) of derivatives, when added to the Feynman
rules will only result in certain
{\em rescaling}
of
$g$'s
as long as one computes the
$S$-matrix.

In the same way we can specify all the 4-leg minimal couplings
contributing at tree level, but in our case it appears to be
unnecessary. The reason is not simple, so let us not discuss their
structure at this stage and suppose that transition to the minimal
parametrization has been done. The main thing one should keep in mind
is that the
$S$-matrix
is completely specified when the values of all the minimal couplings
are given.

The way one assigns certain values to the
$S$-matrix
parameters in perturbation theory is the renormalization prescriptions
(RP's). To obtain our tree level amplitude, we need to specify 3- and
4-leg couplings and masses. Forgetting for a while about 4-leg
couplings we concern ourselves with the remaining parameters. As
pointed out in
\cite{AVVV2},
the resultant parameters are the natural candidates to impose the RP's
under the condition that the renormalization point is taken on shell
and
{\em renormalized perturbation theory}
is used. In this scheme the action is written in terms of
{\em physical}
parameters plus counter terms, the latter are tuned in a way that the
values of those parameters remains unchanged after renormalization.
So, we imply that the Feynman rules are written in the form of
physical part plus counter terms at every loop order and it is the
{\em real parts of physical masses}
that appear in bare propagators. Simply speaking, we impose the
following set of RP's:
\[
\bo{\rm Re}\ V(p_1,p_2,p_3) = G_{phys} {\rm \; at \;}
p_i^2 = M_{i_{phys}}^2,
\]
and
\[
\bo{\rm Re}\ \Sigma(p) = 0 {\rm \; at \; }
p^2 = M_{phys}^2,
\]
for every self-energy
$\Sigma$
and every three-point vertex
$V$.
Now we are at tree level, thus there are no counter terms relevant,
therefore the couplings
$g$
are also physical (experimentally measurable).

There is no phenomenological evidence that the mass spectrum and
spin values of resonances are bounded from above. Therefore we need to
reserve the possibility to work with infinite set of resonances of
arbitrary high spin value. In other words, there is still infinite
number of minimal couplings coming even from three-leg vertices. One
of the main points of our work is that these couplings are
{\em not independent}:
there are
{\em self-consistency conditions}
that restrict their values. We call this conditions as the
{\em bootstrap}
equations.

%%%%%%%%%%%%%%%%%%%%%%%%%%%%%%%%%%%%%%%%%%%%%%%%%%%%%%%%%%%%%%%%%%%%%%
\section{Bootstrap and experimental data}

Because of lack of space we do not discuss here the method of
constructing the well defined expressions for the amplitude at tree
level or at any given order of perturbation theory. It is enough to
say that the main tool allowing to do this is just the celebrated
Cauchy integral formula with the
{\em summability}
and
{\em asymptotic uniformity}
conditions discussed in
\cite{vvv-hsqcd}.
The final expression turns out to be completely parametrized by
the minimal couplings. Moreover, in the case of tree level
$\pi N$ elastic scattering amplitude only triple resultant vertices
enter this expression. The joint contribution of four-leg vertices
turns out to be uniquely determined by masses and triple couplings%
\footnote{
This statement is by no means trivial and requires separate
consideration. The main reason for it is the known values of Regge
intercepts which, by uniformity principle, define the asymptotic
behavior of the tree level amplitude. This analysis will be published
elsewhere.
}.

The bootstrap equations mirror the crossing symmetry of a given order
amplitude within our perturbation scheme. They can be rewritten in a
form of infinite set of numerical equations for the amplitude
parameters
\cite{AVVV1,POMI}.
What is essential to stress here is that the parameters that enter
those equations are all minimal, and hence, as explained in the
previous section, they are physical or (at least, in principle)
{\em measurable}.

Using the renormalized perturbation theory with on-shell RP's at
each loop level one obtains certain set of bootstrap equations which
should be satisfied to ensure self-consistency (usually crossing
symmetry). The form of these equations may vary from level to level,
but
{\em all}
of them are the equations for physical parameters, and the full set of
RP's should be compatible with all of them. To put it another way,
the set of renormalization prescriptions for couplings and masses must
be a solution to the full set of the bootstrap constraints.

We do not know how the solution of this latter set looks like. Even at
tree level their form is highly non-linear. However, if our
perturbation scheme can describe nature, then the experimentally
fitted values of coupling constants and masses must fulfil the system
of bootstrap conditions. That is why we have performed various
calculations to check the consistency of our approach with the
experimental data. Namely, we checked the tree level bootstrap
equations for
$\pi \pi$ and $\pi K$
elastic scattering amplitudes (see
\cite{AVVV1}
and references therein), and recently analogous calculations were
performed for the cases of
$\pi N$
\cite{menu}
and
$K N$
elastic scattering (the latter case is discussed in the talk by
K.~Semenov-Tian-Shansky
\cite{kstsh-hsqcd}).
There were no contradiction found so far, and in most cases examined
the experimental data seem to support our approach nicely.

Apart from the question of formal compatibility with experiment, there
is a question of efficiency. One can ask how many loops should be
taken into account and how many parameters fixed to obtain the
amplitude that could fit well the data at least at some kinematical
region. To check this point we performed a calculation of low energy
coefficients%
\footnote{
Taylor expansion coefficients around the crossing symmetry point.
}
for the
$\pi N$
amplitude. This coefficients measured and fitted in
\cite{Nagels}
are reproduced in our approach with very good accuracy already at tree
level%
\footnote{
Of course, it is partly because this region is relatively far from the
branch cut points. In case if the latter appears close to the
investigated region one should necessary include loops.
},
and to gain reasonable precision it is enough to specify the
parameters of just few lightest resonances. The results of this
analysis were summarized in
\cite{menu};
the details will be published elsewhere.

%%%%%%%%%%%%%%%%%%%%%%%%%%%%%%%%%%%%%%%%%%%%%%%%%%%%%%%%%%%%%%%%%%%%%%
\section*{Acknowledgments}

I am grateful to V.~Cheianov, H.~Nielsen, S.~Paston, J.~Schechter,
K.~Semenov-Tian-Shansky, A.~Vasiliev, V.~Vereshagin and M.~Vyazovsky
for stimulating discussions.

%%%%%%%%%%%%%%%%%%%%%%%%%%%%%%%%%%%%%%%%%%%%%%%%%%%%%%%%%%%%%%%%%%%%%%

%%%%%%%%%%%%%%%%%%%%%%%%%%%%%%%%%%%%%%%%%%%%%%%%%%%%%%%%%%%%%%%%%%%%%%
\end{document}